# Recovering individual emotional states from sparse ratings using collaborative filtering

Eshin Jolly[1*]
Max Farrens[1]
Nathan Greenstein[1]
Hedwig Eisenbarth[2]
Marianne C. Reddan[3]
Eric Andrews[4]
Tor D. Wager[1]
Luke J. Chang[1*]

[1]Department of Psychological and Brain Sciences
Dartmouth College
Hanover, NH, USA

[2]School of Psychology
Victoria University of Wellington
Wellington, New Zealand

[3]Department of Psychology
Stanford University
Palo Alto, CA, USA

[4]Department of Psychology
University of Arizona
Tucson, AZ, USA

Word Count: 7,442
[*]Corresponding Author
eshin.jolly@dartmouth.edu
Computational Social and Affective Neuroscience Laboratory
Department of Psychological Brain Sciences
Dartmouth College
6207 Moore Hall
Hanover, NH 03755
(603) 646-2056

# Abstract

A fundamental challenge in emotion research is measuring feeling states with high granularity and temporal precision without disrupting the emotion generation process. Here we introduce and validate a new approach in which responses are sparsely sampled and the missing data are recovered using a computational technique known as *collaborative filtering* (CF). This approach leverages structured covariation across individual experiences and is available in *Neighbors*, an open-source Python toolbox. We validate our approach across three different experimental contexts by recovering dense individual ratings using only a small subset of the original data. In dataset 1, participants (n=316) separately rated 112 emotional images on 6 different discrete emotions. In dataset 2, participants (n=203) watched 8 short emotionally engaging autobiographical stories while simultaneously providing moment-by-moment ratings of the intensity of their affective experience. In dataset 3, participants (n=60) with distinct social preferences made 76 decisions about how much money to return in a hidden multiplier trust game. Across all experimental contexts, CF was able to accurately recover missing data and importantly outperformed mean and multivariate imputation, particularly in contexts with greater individual variability. This approach will enable new avenues for affective science research by allowing researchers to acquire high dimensional ratings from emotional experiences with minimal disruption to the emotion-generation process.

# 1. Introduction

A fundamental challenge inherent to emotion research is how to measure feelings with high granularity and temporal precision [1]. Unfortunately, there is currently no method that can objectively and reliably track an individual's emotional state [2]. Instead, emotion researchers rely on indirect methods to infer manifestations of these latent states, such as behavior, facial expressions, psychophysiological measures, and self-report. Behavioral measures such as how emotion influences judgment and decision-making [3,4] can provide indirect objective measurements, but are difficult to validate and are frequently uncorrelated with other emotional indicators such as self-reported experience [5]. Facial expressions offer an additional behavioral display of emotional states [6–10], but can be modulated across different communicative contexts [11] and may not directly reflect an internal emotional state [12]. Psychophysiological measures such as electrodermal activity, heart rate variability [13], pupillometry [14], and blood pressure [15] provide continuous measures of autonomic activity that are sensitive to changes in an individual's state of arousal, but cannot directly reveal an individual's nuanced emotional experience [16–20].

Currently, self-report remains the primary method to measure emotional experiences and is associated with the highest overall effect size in emotion elicitations [5] and discriminating between discrete emotional states [20]. This approach is well suited for recording a range of distinct feeling states at discrete points in time [21], and can also be used to measure changes in a single state over time [22]. However, self-report techniques also have significant disadvantages. First, people have a limited capacity to introspect about the processes underlying cognition [23], and some who score high on trait alexithymia are less able to reflect on emotional experiences at all [24]. Second, there are several factors that reduce the reliability of self report techniques. For example, people have difficulty integrating emotional experiences over time and typically report the maximum emotional intensity or the last experienced state, an effect referred to as the *peak-end rule* [25,26]. In addition, self-report measures are typically probed on a cardinal scale, but these judgments are likely reference-dependent and may be better indicated on an ordinal scale [27]. Third, the act of introspecting and reporting likely disrupts the natural emotional response to an eliciting stimulus [1] and the manner in which an emotion is probed (e.g. binary vs continuous) can impact the judgment process [28]. Moreover, it is highly likely that attending to a stimulus while simultaneously monitoring and reporting one's internal state can change the nature of the emotional experience, akin to observer effects in physics, based on our limited cognitive ability to perform dual-tasks [29].

For these reasons, there is a strong need for methodological advances that can improve researchers' ability to measure emotional experiences with high temporal precision and dimensionality but with minimal impact on the emotion-generation process itself. Currently, a popular approach based on latent factor analyses is to use a low dimensional embedding of emotional experiences such as valence and arousal [30–32]. While this approach dramatically simplifies the collection of emotional responses and has been shown to be highly reliable, it does not reflect the richness of emotional experiences [33]. This raises the possibility that affective science itself unwittingly becomes limited to the study of low dimensional explanations by undersampling the diverse range of psychological experiences of emotion, i.e. reflecting a Flatland Fallacy [34,35]. Indeed, using richer and more diverse stimulus sets than static pictures such as the International Affective Picture System [36] has revealed a substantially higher dimensional space of emotion experiences than can be captured by low dimensional embeddings [37].

In this paper, we propose and test a novel solution for measuring self-reported emotional states and affect-influenced behavior using a high-dimensional emotional space with high temporal precision. The basic idea is to *sparsely* sample an individual's feeling states across many different dimensions and/or time-points, and then infer the unreported feeling states using computational modeling. In principle, this would allow researchers to increase both temporal precision and the dimensionality of emotional states without dramatically increasing the participant's workload, thereby minimizing the impact of the reporting process on the emotion-generation process itself. Central to our approach is a machine-learning technique known as *collaborative filtering* (CF), which is commonly employed in recommender systems that power commercial products such as Netflix, Amazon, and Spotify, where the goal is to infer an individual's preferences on many items (e.g. movies, products, and music) when their preferences are only known for a few items [38]. CF makes predictions about an individual's *unobserved* responses to a set of items by leveraging the *similarity* between that individual and other individuals' *observed* responses to different items [39]. For example, knowing that person A rates a set of items similarly to persons B, C, and D, CF can infer person A's *unobserved* ratings on a different set of items using a combination of the responses made by persons B, C, and D on those items. Importantly, this technique makes no assumptions about the underlying distribution of ratings across participants and permits variations in individual differences in ratings compared to simply imputing the sample norm.

Here, we introduce *Neighbors* (https://neighbors.cosanlab.com), an open source Python toolbox for performing CF, and validate its ability to accurately infer participants' emotional responses across three distinct prototypical experimental contexts (Fig 1). First, we utilized affective images to measure how well responses were recovered when evoked by static stimuli that have been normed to minimize individual variability [36,40]. We expected CF to provide only marginal benefits over simple mean imputation given the explicit minimization of individual variation in normative ratings of these stimuli. Second, we utilized emotional videos to assess how well CF could recover *time-series* measurements of the valence of an experience unfolding over time. Because there is greater individual variability in the dynamics of participants' emotional experiences in this experimental context, we anticipated CF would provide substantial benefits over mean imputation. Lastly, we utilized previously published behavioral data to evaluate how well CF can recover individual behavioral *decisions* with known sub-groups of people that vary in their motivations and preferences (e.g., guilt-aversion or inequity-aversion) [41]. Given that each sub-group varied in size and behavior, this allowed us to test whether different CF approaches behaved differently for each sub-group of individuals *without* providing sub-group information to any algorithm.

Our general approach consisted of collecting dense responses to each stimulus type and then validating CF inference by randomly masking out a subset of these responses to generate sparse versions of the same data. We compared the performance of two families of CF approaches: neighborhood methods implemented via K-Nearest Neighbors and latent factor approaches implemented via non-negative matrix factorization [38]. Neighborhood methods make explicit use of the similarity between individuals to make predictions via weighted averaging, while latent factor methods instead decompose observed responses to learn latent spaces of individuals, which are then combined to make predictions for specific individual-item responses. We compared the performance of these two models across datasets to two different control models. As a baseline, we computed the average ratings across all participants. We also compared our models to a state of the art missing data imputation technique called multivariate imputation with chained equations (MICE) commonly used in psychological and affective science [42].

## 2. Methods

### 2.1 Datasets

All reported analyses are based on three datasets. Datasets 1 and 2 were specifically collected for the purposes of this manuscript, while Dataset 3 has been previously published [41]. All participants provided informed consent and experimental procedures were approved by the Institutional Review Boards at Dartmouth College (dataset 3 was approved by the CMO Arnhem–Nijmegen, the Netherlands).

#### 2.1.1 Affective Images
Dataset 1 was used to assess the ability of CF to recover multiple dimensions of individual emotion ratings from static images. Participants ($N$ = 316) were recruited from Amazon Mechanical Turk (AMT) and paid $1 for approximately 12 minutes of their time ($M$ = 11.97 +/- 8.15 minutes). All participants were fluent English speakers over 18 years old, but no additional demographic data was collected. Each participant rated how 112 emotionally evocative images made them feel along one of six basic emotion ratings (i.e., anger, sadness, joy, surprise, fear, and disgust), using a 100-point visual analog scale (VAS) initialized at the midpoint at the start of each trial [43]. Images were selected from the IAPS stimulus set [36] and the Geneva Affective Picture Database (GAPED) [40]. Each participant rated *all* images for a single emotion dimension resulting in a minimum of 52 unique ratings per emotion-image combination: $N_{Anger}$ = 53; $N_{Fear}$ = 53; $N_{Joy}$ = 53; $N_{Surprise}$ = 53; $N_{Sadness}$ = 52; $N_{Disgust}$ = 52. The experimental task was developed using the psiTurk framework [44] and deployed on a custom server hosted on OpenShift (https://www.openshift.com/).

#### 2.1.2 Emotional Videos
Dataset 2 was used to assess the ability of CF to recover continuous emotion time-series ratings in response to a dynamic stimulus. Participants ($N$ = 494) were recruited from AMT and continuously rated how positive or negative they felt on a bipolar 100 point VAS while watching one of 8 emotionally arousing video clips that varied in duration ($M$ = 10.33 minutes; $SD$ = 4.04 minutes; *Max*: 17.56; *Min*: 5.5). Participants were fluent English speakers over 18 years old with a 98% or higher Mturk approval rating and experience completing at least 100 Assignments.

Ratings were collected at a temporal resolution of 1Hz. All clips were sourced from The Moth Radio Hour (https://themoth.org/), a program consisting of live performances of emotionally-moving personal stories. To fairly compensate participants for their time, three separate HITs were launched on AMT for short, medium, and long videos that compensated at $1.50, $2.50, and $3.50 respectively. The experimental task was developed as a general purpose web application for continuous measurement using the SvelteJS (https://svelte.dev) framework [45] and hosted on Netlify (https://netlify.app/).

Several sources of participant attrition occurred during data collection. Participants were excluded (N=249) because they: (a) failed a custom captcha-like bot-check that asked them to solve an arithmetic question, (b) stopped participating before making any ratings, or (c) watched the video in its entirety but failed to make any ratings. An additional 42 participants were excluded due to programmatic differences across two batches of data collection that resulted in poor quality data. In batch 1, participants were provided "jump keys" – keys that allowed them to quickly change their ratings to 0 or 100 from anywhere in the rating box. Additionally, in this same batch, a technical glitch did not reset an input acceleration factor between trials, which potentially allowed participants' ratings to oscillate rapidly. To address this, participants whose rating time-course "jumped" more than 20 times were excluded from all analyses. The removal of jump keys and a corrected acceleration factor were employed in batch 2. Altogether, this left a total of 203 participants whose data were analyzed for this manuscript. Each participant watched and rated at least one video clip in its entirety while a smaller number of participants watched and rated two ($N = 34$) or three video clips ($N =13$). This resulted in a minimum of 28 unique ratings per video ($N_{Video} = 28\text{-}46$; *median* = 31).

### 2.1.3 Social Decisions

Dataset 3 was used to assess the ability of collaborative filtering to recover interpersonal decisions guided by social emotions such as guilt, fairness, and moral opportunism [41]. While we refer readers to the original published paper for more details, in brief: participants (N=66) were recruited from the Nijmegen student population at the Donders Institute for Brain, Cognition, and Behavior to play a variant of an Investment game [46] called the Hidden Multiplier Trust Game (HMTG) [47]. Across 80 experimental trials, participants ("trustees") were shown financial investments made by previous participants ("investors") [48] and tasked with deciding how much of this investment to return and how much to keep for themselves. Investors believed that their investment decisions would be multiplied by a fixed amount (x4), but only trustees observed the true multiplier (2x, 4x, 6x). The resulting information asymmetry combined with a computational model, allowed the original authors to probe individual phenotypic variation in participant's social preferences underlying their motivation to reciprocate in the game (e.g. guilt-aversion, inequity-aversion, greed, and moral-opportunism). Only participants with a complete set of 80 decisions were analyzed for this manuscript and trials in which the original investment amount was 0 were excluded. In addition, we averaged each participant's decisions to trials with the same investment-multiplier combination. This left a total of 60 participants with 76 decisions each.

## 2.2 Approach and Algorithms

For ease of understanding and consistency with extant literature, throughout this manuscript we use the terminology *users* as individual participants in each dataset, *items* as images (Dataset 1), time-points (Dataset 2) or experimental trials (Dataset 3), and *ratings* as the subjective feeling judgments (Datasets 1 & 2) or decisions (Dataset 3) that each user made about each item. All datasets comprised dense ratings, i.e. each user-item rating was observed. However, in real world datasets and many experimental settings, obtaining dense ratings for a large set of users and/or items is infeasible, resulting in sparse data, i.e. not all users have rated all items. Therefore, to test how well CF could be used to make accurate predictions for missing data, varying degrees of sparsity were imposed on each dataset via masking random data points in 10 percent increments. This procedure simulates scenarios in which each user only rated a subset of all items, and allowed us to evaluate how well the CF was able to recover the missing ratings across varying levels of sparsity (see 2.2.8 for details).

### 2.2.1 *Overview*

Collaborative filtering (CF) is a machine-learning technique that falls into a broader class of approaches for building recommender systems [38]. Broadly, the goal is to make *predictions* (recommendations) about a user's preferences or behavior (e.g., ratings) based on sparsely sampled data reflecting their actual preferences or behavior. The key insight behind CF is leveraging the *similarity* between users' ratings to inform these predictions. CF algorithms broadly fall into two categories [39]: (1) memory-based models such as neighborhood techniques and (2) model-based techniques such as latent factor models. Neighborhood techniques, such as K-Nearest Neighbors (KNN), combine scores from a user's "neighbors" to make predictions for that user. For example, if we assume that each person's preferences are similar to their friends, we could infer an individual's preference for an item by averaging their friends' preferences. This is conceptually similar to a clustering analysis, where users are clustered based on their position within an item embedding space. While clustering algorithms attempt to group participants into discrete groups, memory-based CF techniques attempt to make predictions about a user's preference for an item, based on averaging responses from other participants that rated other items similarly. In contrast, model-based techniques such as non-negative matrix factorization (NNMF; [49]) attempt to learn latent dimensions which characterize variation across users. This method is more conceptually similar to decomposition analyses such as principal components analysis (PCA) or factor analysis, which are often used to uncover the latent structure of how items relate to each other in order to compress or simplify an item embedding space. However, rather than learning the latent structure of items, CF instead learns the latent structure of how users relate to each other. These latent representations are then used to make predictions about how a user might rate a particular item. It is important to note that while CF techniques share conceptual similarity to familiar analytic techniques, they are able to be estimated from *sparse data*, which would be problematic for more traditional clustering or latent factor models.

In the present work, we computed predictions using four different models: (1) A baseline model using simple mean imputation by averaging across all users' ratings for a specific item to generate predictions for a single unobserved user's rating for that item; (2) MICE [42] - a state of the art missing data imputation technique; (3) a KNN neighborhood model, i.e. a user-similarity weighted-mean model restricted to the 10 most similar users to a single unobserved user; and (4) a NNMF model trained using stochastic gradient descent (NNMF sgd). In all cases, each algorithm received a two-dimensional *users* x *items* ratings matrix $V$ where a single row $i$ and column $j$ index indicated a specific user's rating of a specific item. Each algorithm was trained using observed ratings to predict unobserved (masked out) ratings and performance was assessed by comparing model predictions to the true ratings.

### 2.2.2 *Mean (baseline)*

The mean model was employed to serve as a baseline comparison for other approaches. The heuristic was straightforward: to make a prediction about a specific item for any unobserved user, we simply used the *mean* of all observed users' ratings for that same item.

### 2.2.3 *Multivariate Imputation with Chained Equations (MICE)*

The MICE approach works by first performing an initial imputation of missing values using the mean of each column of the user x item matrix, i.e. the mean of observed ratings for each item. It then estimates a separate model (e.g. linear regression) for each column of this matrix (each *item*) by using all *remaining* columns (all other *items*) as predictors. We employed MICE as implemented in scikit-learn [50] using the IterativeImputer class with a linear regression estimator. Ten imputation rounds were performed using this approach in a random-order across all items and the output from the final imputation was used as predictions for missing ratings, constrained to the range of the data.

### 2.2.4 *K-Nearest Neighbors*

The KNN approach to CF relies on a representation of the similarity between all users' calculated from their observed ratings across all items. Such techniques assume that there will be a variety of response profiles which will be shared by many participants and can be leveraged to make predictions about unobserved ratings. The KNN approach specifically uses a subset of nearest neighbors $N$ for a given user $u$ based on their similarity. To generate predictions for a specific item $i$, ratings from each neighbor $v$ for that item $r_{v,i}$ are weighted by this similarity and aggregated to make predictions for a given user's unobserved rating for that same item $\hat{r}_{u,i}$ [38]. Formally, this can be described as

$$\hat{r}_{u,i} = \frac{\sum_{v \in N_{j=1}^k(u)} r_{v,i} \cdot sim(u,v)}{\sum_{v \in N_{j=1}^k(u)} sim(u,v)}$$

(Eq. 1)

For the reported analyses, we used Pearson correlation to compute user similarity and limited the number of neighbors to $k = 10$, but KNN works with any distance/similarity metric and our Neighbors software toolbox supports several common metrics including cosine similarity and rank metrics such as spearman's rho and kendall's tau. An important limitation of KNN approaches occurs for highly sparse data where $N$ is fewer than the expected number of neighbors (or 0) or neighbor ratings for a specific item, e.g. $r_{v,i}$ are also unobserved. This occurs because KNN typically only makes use of neighbors for which similarity is positive and in extremely sparse data situations it is possible for no positive similarity scores to exist. Therefore in practice $N$ reflects the *maximum* number of neighbors used to make a prediction. When $N$ is 0, our implementation utilizes the mean for all observed users' ratings for item $i$. In cases where all other users' ratings for item $i$ are unobserved, we utilize the global user-item mean across all observed ratings.

### 2.2.5 *Non-Negative Matrix Factorization*

Non-negative matrix factorization (NNMF) is a model-based CF technique that differs from neighborhood based algorithms by learning latent factors (a model) that capture variation in users and items [38]. Closely related to singular value decomposition (SVD), NNMF aims to factorize a matrix $V$ into two non-negative matrices $W$ and $H$, such that $V \approx WH$. Here, $V$ is a $u \times i$ matrix which represents how each user $u$ rated each item $i$, $W$ is a $u \times f$ matrix where $f$ represents the number of latent factors and each value represents the affinity or weight for each user $u$ on each latent factor $f$, and $H$ is a $f \times i$ matrix which represents the affinity or weight for each item $i$ on each latent factor $f$. These factors can then be combined to make predictions for any user-item rating according to the formula:

$$\hat{r}_{u,i} = q_i^T p_u$$

(Eq. 2)

where $q_i$ is a vector in $H$ representing factor scores for item $i$ and $p_u$ is a vector in $W$, representing factor scores for user $u$ [49].

Matrix factorization techniques have the added flexibility of using low-rank approximations, whereby $f < u$ or $f < i$. This has the effect of compressing or denoising the learned model by discarding potentially uninformative user or item factor loadings when making predictions [39]. Because datasets making use of CF are often sparse, conventional efficient techniques for computing matrix factors (e.g. SVD) are undefined when working with missing data [51]. For this reason, most implementations instead employ numerical optimization techniques to learn these latent factors in an iterative fashion. In this paper we use *stochastic gradient descent* (SGD) to learn these latent factors based solely on the observed ratings. We additionally trained a NNMF model using the multiplicative updating rule proposed by Lee and Seung [52] however this algorithm does not natively support sparse matrix updates and our implementation performs worse than even the baseline mean model (see Supplementary Materials).

*Stochastic Gradient Descent*

In the present work, we employed a flexible version of NNMF trained via SGD that includes additional parameters that capture variation in users or items that are independent from their interaction [38]. These parameters, referred to as *biases,* capture the tendency for some users to provide higher or lower ratings on average compared to other users, and likewise for some items to receive higher or lower ratings on average compared to other items. Predictions for a specific user-item rating $\hat{r}_{u,i}$ are made according to the formula:

$$\hat{r}_{u,i} = \mu + b_u + b_i + q_i^T p_u$$

(Eq. 3)

where $\mu$ is the global mean of observed user-item ratings, $b_u$ is user $u$'s bias, $b_i$ is item $i$'s bias and $q_i^T p_u$ is the interaction between factor scores for item $i$ and factor scores for user $u$.

For training, we initialized user and item biases as zero vectors and user and item factors as random normal matrices scaled by the number of factors. Each iteration began with randomly shuffling the observed ratings and subsequently making predictions while individually updating each user-item error by minimizing the following regularized loss function:

$$L = \sum_{r_{ui} \in R_{\text{observed}}} (r_{ui} - \hat{r}_{ui})^2 + \lambda_{bi} b_i^2 + \lambda_{bu} b_u^2 + \lambda_{qi} ||q_i||^2 + \lambda_{pu} ||p_u||^2$$

(Eq. 4)

Each $L_2$ regularization term $\lambda$ is a hyper-parameter that prevents overfitting of item biases, user biases, item factors, and user factors, respectively. Our implementation allows each regularization term to be set independently, but many popular implementations utilize a single $\lambda$ for all model parameters often learned via nested cross-validation [38]. For each prediction $\hat{r}_{u,i}$, a separate prediction error $e_{u,i}$ is calculated by comparing predicted ratings to true ratings $r_{u,i} - \hat{r}_{u,i}$. Each error is used to update model parameters using the following update rules:

$$b_i \leftarrow b_i + \gamma(e_{u,i} - \lambda_{bi} b_i)$$

(Eq. 5)

$$b_u \leftarrow b_u + \gamma(e_{u,i} - \lambda_{bi} b_u)$$

(Eq. 6)

$$q_i \leftarrow q_i + \gamma(e_{u,i} \cdot p_u - \lambda_{qi} q_i)$$

(Eq. 7)

$$p_u \leftarrow p_u + \gamma(e_{u,i} \cdot q_i - \lambda_{pu} p_u)$$

(Eq. 8)

where $\gamma$ is a learning rate that determines the size of the update. Here, we use a single learning rate for all parameter updates, but alternate implementations can use independent learning rates for each parameter to further improve model fitting [38]. A single training iteration thus comprises prediction and error updates utilizing *all* observed ratings (subject to the constraint that user and item factors matrices are non-negative).

For all reported analyses, we set $\gamma = .001$, $f = min(N_{users}, N_{items})$, and used no regularization on other model parameters ($\lambda_{all} = 0$). While popular datasets like the Netflix challenge can make use of fixed hyper-parameter values established via consensus [38,53], we unfortunately did not have enough data to simultaneously tune these hyperparameters and evaluate the accuracy of the models in this study. Future work might explore tuning these hyperparameters which are likely to be influenced by specific features of the dataset such as sparsity, number of users, number of items, user similarity, etc. To balance each model's training time with overall computation time, training ran until changes in the error were <= 1e-6 or 1000 iterations were reached.

### 2.2.6 *Kernelized Time-Series Predictions*

In the unique case of Dataset 2, observations consisted of time-series data in which users provided one rating per second of each video. Time-series data often demonstrate intrinsic auto-correlation, such that ratings at points closer in time are likely to be more similar than ratings at points more distant in time. We exploited this structure by convolving the observed data with a kernel prior to model fitting. This data augmentation strategy effectively "dilates" each observed rating by filling-in surrounding sparse timepoints with the same or a weighted-average rating. To evaluate this approach, we fit several additional models at each level of sparsity using a box-car kernel of varying width: 5s, 20s, or 60s. This procedure reduces the overall sparsity of the data via locally-weighted mean imputation. Critically, this procedure occurs *after* dataset masking, so that dilated values only make use of observed ratings with no "leakage" of unobserved ratings used for model evaluation [54]. While the kernel shape and widths were chosen arbitrarily, in principle a kernel of any shape (e.g. gaussian) could be treated as an additional model hyper-parameter and thus further tuned via nested cross-validation [55].

### 2.2.7 *Cross-Validation and Model Evaluation*

To take advantage of the fully observed (dense) ratings in each dataset, we employed randomly shuffled cross-validation to generate sparse versions of each dataset for training and tested performance on left-out (unobserved) ratings. Specifically, 10-90% of the total number of ratings in each dataset were masked out as "missing" values in each user x item ratings matrix. Each algorithm used this sparse matrix of observed ratings to predict values for each "missing" rating, thereby completing the full user x item rating matrix with predicted values. These predictions were compared to the true values of the missing ratings using Root-Mean-Squared-Error (RMSE), which can be interpreted as error on the original scale of the data (see Fig 2 for an illustrative example). To facilitate comparisons across datasets which utilized different ratings scales, we computed normalized error as:

$$NormalizedError = \frac{RMSE}{maxR - minR}$$

(Eq. 9)

This procedure bounds all model error scores to the range [0,1] with values near 0 reflecting perfect performance (no deviance from true ratings) and values near 1 reflecting maximal possible error (largest possible deviance from true ratings)[1] [56]. Scores can be interpreted as the percentage inaccuracy relative to the scale of the ratings. For each level of sparsity, this train/test procedure was repeated for 10 iterations using a new random mask for each iteration. Because psychological researchers often focus on generalizing observations across individuals [57], error was calculated *separately* for each user (by averaging normalized error over iterations) and subsequently averaged. This form of aggregation better captures user-level clustering in the data, which can change the outlook of a model's performance if it fails to consistently make accurate predictions for the majority of users. This differs from conventions in the machine learning literature which typically aggregate performance across *all* ratings without first aggregating by user, and tends to produce higher performance estimates [38]. This estimation was performed separately for each emotion type in Dataset 1 and for each video in Dataset 2. However, because results did not differ substantially across different emotions or videos, all reported results reflect aggregated performance over emotions (Dataset 1; Fig S1) or videos (Dataset 2; Fig S2).

### 2.2.8 Model Comparison

To compare algorithm performance we fit a series of multilevel regression models using Pymer4 [58]. For overall comparisons in each dataset, we predicted normalized error using algorithm, sparsity level, and their interaction as predictors. We dummy-coded algorithm type and sparsity level as categorical predictors, with the Mean model and 10% of observed data as the reference group for each predictor respectively. Using least-squares mean estimation [59], we performed inference on regression coefficients to compare how much the KNN and NNMF model performance exceeded the Mean model at each level of sparsity and how well MICE compared to each CF algorithm. To account for the fact that the same user ratings were utilized in each dataset, random intercepts were included for each user; random slopes were not fit as they were either inestimable (the number of random-effect parameters exceeded the number of observations) or because models failed to converge when included. Inference was performed using p-values computed using Satterthwaite approximated degrees of freedom with Bonferroni correction for multiple comparisons [60].

For the kernelized dilation analysis on Dataset 2 we modified the overall regression model by reducing the number of dataset sparsity levels to three (10%, 50%, 90%) to aid in interpretability and dummy-coded dilation (with no dilation as the reference group) and its interaction with other model terms as additional categorical predictors. Least-squares mean estimation was used to compare how each level of dilation changed a single model's performance at each level of sparsity against that model's undilated performance at the same level of sparsity. For the phenotype analysis on Dataset 3, we extended the overall regression model by including phenotype and its interaction with other model terms as additional categorical predictors. Least-squares mean estimation was used to compare the KNN and NNMF models to the Mean model and the MICE model to each CF algorithm at each level of sparsity within a single phenotype. In both cases reported p-values were adjusted using Bonferroni correction [59].

---

[1] In rare cases it is possible for normalized error to exceed 1 when model predictions are especially inaccurate and outside the range of the original rating scale.

## 2.3 Python Toolbox

All algorithms tested in this paper were implemented using a custom, open-source Python toolbox available on GitHub and PyPi called Neighbors ([http://neighbors.cosanlab.com/](http://neighbors.cosanlab.com/)). The toolbox's API is inspired by popular machine-learning toolboxes such as scikit-learn [61]. The toolbox is designed with researchers used to working with repeated-measures data in mind, in which multiple observations come from the same individual. Researchers can supply a long-form "tidy" dataframe where each row consists of a single rating for a specific user-by-item combination [62], and neighbors will transform this into a suitable (possibly sparse) user-by-item matrix for model training. The toolbox provides additional helper methods, e.g. to build summaries of model fits, produce plots, and estimate performance in a cross-validated function with optional parallelization. We have optimized some of the more computationally demanding linear-algebra routines using just-in-time (JIT) compilation via Numba [63]. Unlike related toolboxes, neighbors supports working with time-series data via temporal up and downsampling and applying kernels to "dilate" sparse ratings. By making neighbors freely available, we hope to encourage other researchers to take advantage of collaborative filtering when working with social and emotion data and to contribute their own enhancements.

# 3. Results

## 3.1 Inter-individual dataset variability

Before performing any collaborative filtering analyses, we were first interested in assessing the overall inter-individual variability in each of the three datasets (Fig 1) as we expected CF to provide few if any benefits over simple mean imputation when variability was low. To do so, we computed the pairwise intersubject pearson-correlation (ISC) [64] separately for each dataset. The mean and standard deviation of the ISC in each dataset allowed us to summarize the similarity of participants' ratings, such that higher mean ISC with lower standard deviation reflects more consensus amongst participants while lower mean ISC and higher standard deviation reflects less consensus amongst participants. Overall, we observed the highest rating similarity for affective images: $ISC_{Mean}$ = 0.72, $ISC_{SD}$ = 0.19. Emotional videos demonstrated much less rating consensus and more variability across participants: $ISC_{Mean}$ = 0.20, $ISC_{SD}$ =

0.39 Social decisions were also more idiosyncratic across individuals: $ISC_{Mean}$ = 0.34, $ISC_{SD}$ = 0.32. These results demonstrate that the datasets differ substantially in how consistently participants responded to each of these different tasks. Given this variability, we expected collaborative filtering to best outperform mean ratings when participants had more idiosyncratic responses to the task.

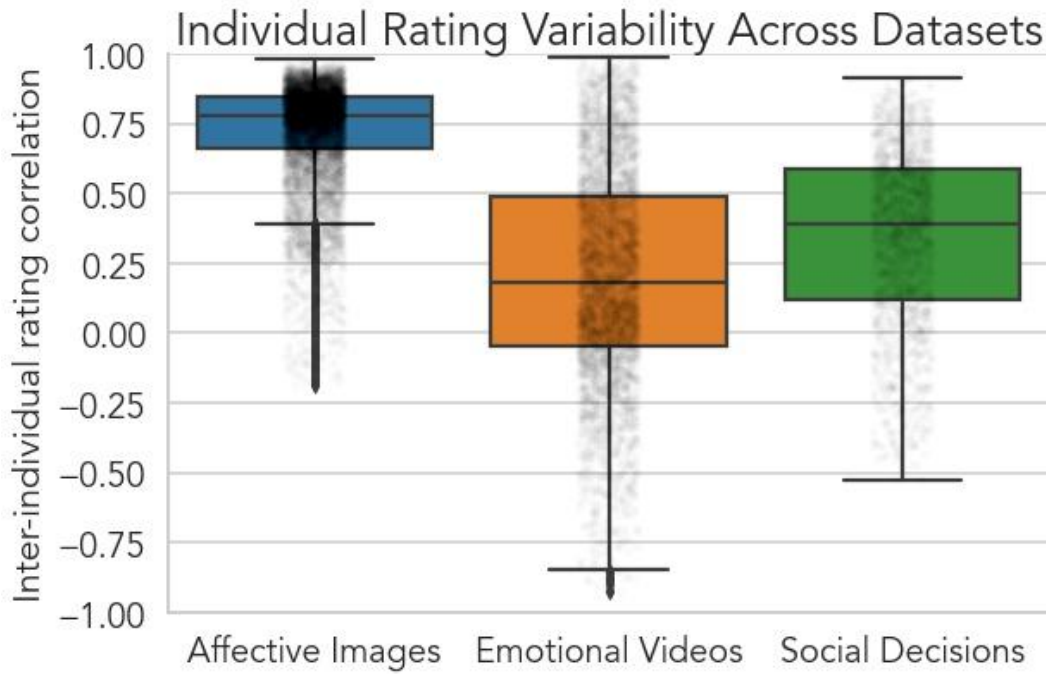

**Figure 1 | Inter-individual rating variability for each reported dataset**  Each analyzed dataset reflected prototypical experimental contexts studied in affective science: static ratings of affective images (blue); time-series rating of emotions videos (orange) and behavioral decisions in a previously published study (green). Critically, each dataset also reflected varying amounts of inter-individual variance depicted here as the average pairwise correlation between individuals' ratings in a given dataset (y-axis). Higher correlations reflect more similar ratings across pairs of individuals and thus *lower* inter-individual variability, while correlations closer to 0 reflect less similar ratings across pairs of individuals and thus *high* inter-individual variability. Because affective images are normed stimuli, they elicit very little inter-individual variability by design, whereas dynamic time-series ratings of emotional videos elicit the highest inter-individual variability. Social decisions also exhibit high inter-individual variability albeit less than emotional videos due to known *sub-groups* of individuals who make similar decisions.

## 3.2 Recovering Single Affect Ratings

Next, we evaluated how well CF could recover missing emotion rating data elicited using a standard static image elicitation paradigm in Dataset 1. In this dataset, participants viewed emotionally evocative images selected from the IAPS and GAPED picture sets and rated the intensity of their emotional response on a single discrete emotion dimension. Because the stimuli are static and provide minimal contextual information and have already been extensively normed, we anticipated low levels of individual variability [36,40] and expected that CF should perform at least as well as mean imputation.

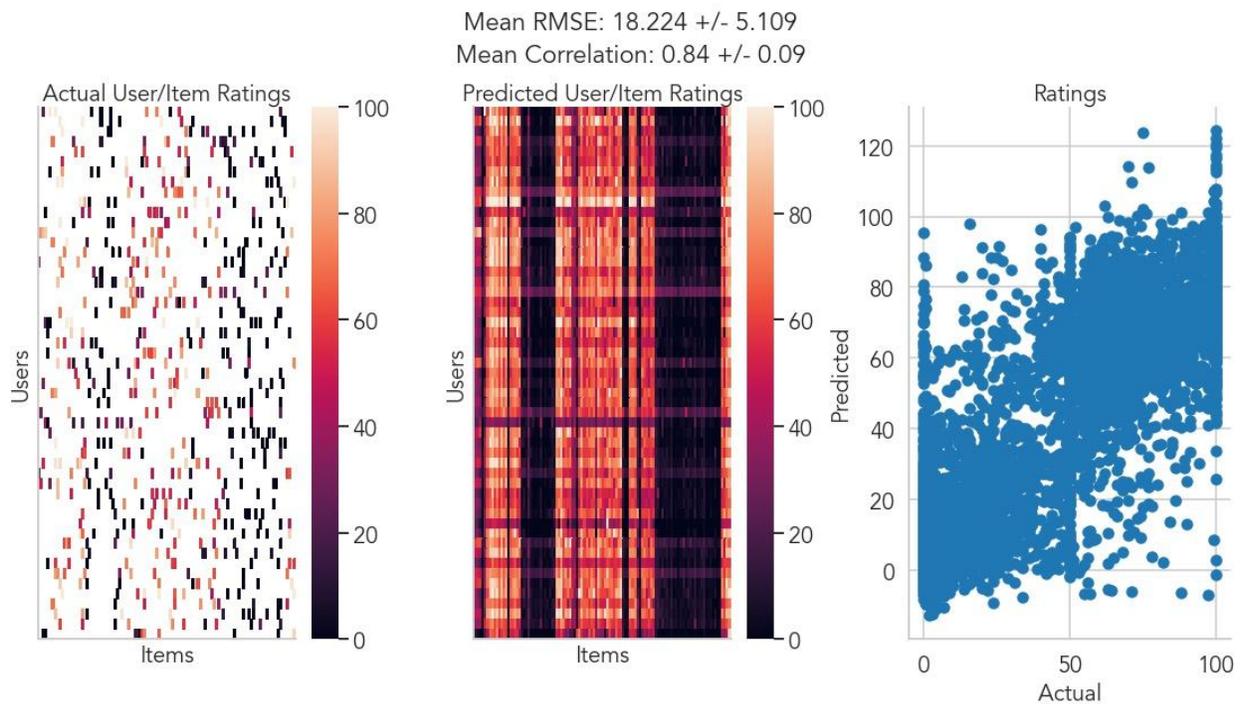

**Figure 2 | Illustrative NNMF SGD model predictions for single emotion dimension with 90% sparsity.** An example of training a NNMF collaborative filtering model via stochastic gradient descent on a single emotion dimension (joy) from the affective images dataset. **Left**: True user x item rating matrix after masking out 90% of observed ratings. **Middle**: Predicted user x item rating matrix after training NNMF model with stochastic gradient descent; **Right**: Scatter plot comparing model predictions for unobserved ratings to true ratings made by each user. Because model predictions are not bounded, in some cases poor predictions extend beyond the range of the data (i.e. less than 0 or greater than 100) Mean RMSE reflects the model error on the original scale of the data calculated separately per used and averaged. Mean correlation reflects the correlation between model predictions and a user's true ratings calculated separately per user and averaged.

Consistent with our predictions, CF only provided marginal benefits for recovering ratings (Fig 3). When less than 40% of ratings were observed, NNMF models had significantly less error in rating recovery compared to the Mean model (all ps < .01), but reduction in error was small, e.g. 10% data $b$ = 1.5% [1% 1.9%][2] improvement. In the same range of sparsity levels, KNN models performed significantly worse than the Mean model (all ps < .01), but these differences were also small, e.g. 10% data $b$ = 2.8% [2.3% 3.2%] more error. All algorithms performed similarly with a moderate amount of data (50-60%). As the amount of data increased, the NNMF model showed slightly worse performance than the Mean, e.g. 90% data $b$ = 2.0% [1.6% 2.5%], while the KNN model showed slightly better performance than the Mean, e.g. 90% data $b$ = 1.0% [0.5% 1.4%]. The MICE model performed significantly worse than all other models, including the

---

[2] 95% confidence intervals assuming a quadratic log-likelihood surface (Wald method)

Mean at all levels of sparsity (all $p$s < .01) except 20% where it performed similarly to the KNN model ($p$ = 0.420). Additionally about 45% of MICE estimations failed when only 10% of the data were observed (Fig S3). Altogether, these results suggest that for datasets in which the measurement context is not expected to induce variability in user ratings (e.g. normed static images) CF may provide only marginal benefits (Fig 3), and simpler techniques such as mean imputation are sufficient for predicting unobserved ratings.

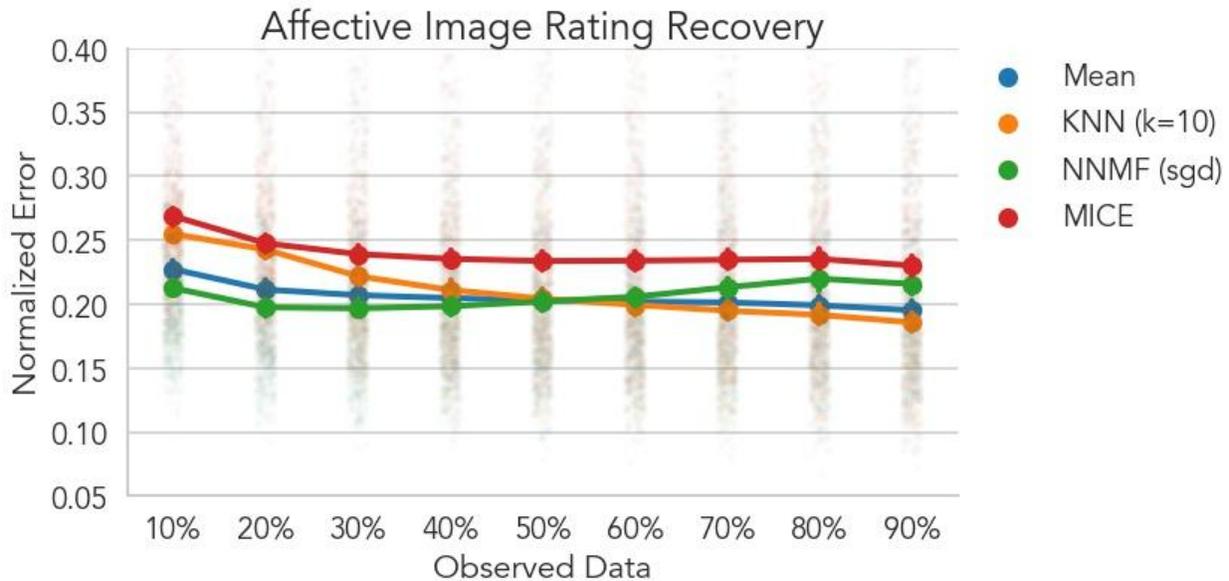

**Figure 3 | Affective image rating recovery.** Each line depicts the cross-validated normalized RMSE averaged across users and emotions. Error bars depict 95% bootstrapped confidence intervals across users and faded points depict mean error for each user over cross-validation folds. Lower values indicate better model performance and can be interpreted as the percentage error on the original scale of the data. MICE models performed similarly or worse than all other models at every level of sparsity. While significant differences between both CF algorithms and the baseline Mean model were observed when data were highly sparse or dense, the magnitudes of these differences were extremely small (~1-3% improvement or decrement). This is likely because the IAPS and GAPES image sets used to evoke responses in these data were specifically constructed to maximize user agreement based on normative ratings. CF may therefore offer limited benefit over simpler techniques like mean imputation for similar data.

## 3.3 Recovering Time-Series Emotion Ratings

Next, we examined how well CF was able to recover continuous affective ratings in response to dynamic naturalistic stimuli. In this experiment, participants watched 8 different videos selected from the Moth Radio Hour, in which a speaker tells an engaging autobiographical story, and simultaneously continuously rated the intensity of the affective experience on a bipolar rating scale (positive-negative). Based on our previous work with naturalistic stimuli [11,65], we anticipated much greater individual variation in emotional experiences as appraisal processes are likely influenced by participants' previous life experiences, current homeostatic states, and future goals. Overall, we found that CF, and NNMF in particular, provided substantial benefits in recovering ratings at all levels of sparsity compared to mean imputation. When only 10% of ratings were observed, NNMF models with no dilation (Fig 4 green line) had an approximately 10% reduction in error rating recovery compared to the Mean model $b$ = 9.9% [8.9% 10.9%], $p$ < .001. This pattern was true for all additional levels of sparsity with model improvement ranging between 10-13% (all $p$s < .001). NNMF with a dilation kernel of 5s performed even better (Fig 4 purple line), with reduced error rates between 15-20% (all $p$s < .001). KNN models only exceeded the Mean model when at least 30% of data were observed, with error reduction ranging between 2.5-5.7% (all $p$s < .001). In addition, as the amount of observed data increased, Mean model error rates quickly converged but never improved beyond 25% error, suggesting that the average rating for each item (time-point) served as a poor model for individual variability in each user's ratings.

The MICE model performed significantly better at all levels of sparsity compared to both the Mean and KNN models (all $p$s < .01). Compared to the NNMF model with no dilation, MICE performed significantly worse when 20% or less data was observed (all $p$s < .01), similarly when 30-40% of the data were observed, and significantly better when 50% or more data were observed. Because missing items in this dataset were time-points, the high performance of MICE was likely due to inherently leveraging the autocorrelation and covariance across time-points in the dataset. However, when even a small amount of temporal dilation was used, NNMF significantly outperformed MICE at all levels of sparsity (all $p$s < .05). Additionally, when 30% or less data was observed, MICE began to fail to impute data because too few observations were available, and at 10% observed data was unable to impute ratings at all (Fig S3).

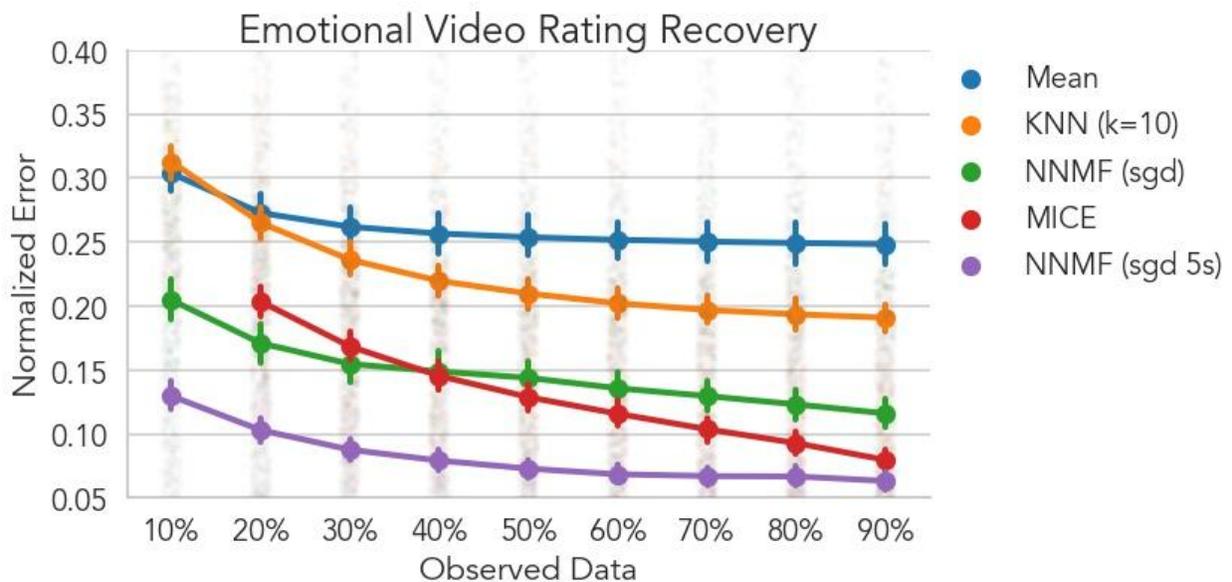

**Figure 4 | Emotional video rating recovery.** Each line depicts the cross-validated normalized RMSE averaged across users and videos. The NNMF algorithm was additionally estimated with a 5s dilation kernel in this comparison (purple line). Error bars depict 95% bootstrapped confidence intervals over users and faded points depict mean error for each user. Lower values indicate better model performance and can be interpreted as the percentage error on the original scale of the data. NNMF with no dilation (green) significantly outperformed the Mean and KNN model at every level of dataset sparsity with average error reduction ranging from 10-13%. NNMF with 5s of temporal dilation (purple) had the lowest overall error rates across all levels of sparsity, outperforming the Mean, KNN and MICE models. KNN performance only exceeded the Mean model when at least 30% of data were observed with error reduction ranging between 2.5-5.7%, likely due to issues computing reliable similarity estimates between users when data were very sparse. When data were especially sparse, MICE performed worse than NNMF without dilation, but performed similarly or slightly better when more data were observed. Additionally MICE estimation for these data were not possible when sparsity was extremely high (10%). For dynamic time-series data with lots of individual variation, CF with temporal dilation offers substantial benefits over simpler techniques like mean imputation.

We also performed additional exploratory analyses to examine how dilation kernel sizes interacted with CF algorithms for rating recovery recovery of emotional videos. In particular, when only 10% of ratings were observed, any amount of dilation reduced model error for both algorithms including the Mean model, all $p$s < .05 (Fig 5). Maximal improvement compared to no dilation at this level of sparsity ranged from 5% (Mean model) to 7-10% (KNN and NMMF). When 50% of ratings were observed, 5 or 20s of dilation further improved performance for KNN and NNMF models, but not the Mean model, all $p$s < .05. Peak performance at this level of sparsity was exhibited by NNMF with 5s dilation which recovered ratings with only 7.1% [5.8% 8.3%] error, $p$ < .001 (Fig 5 right panel, middle group, orange bar). However, at this level of sparsity, the effect of *over*-dilating became apparent: at 60s of dilation KNN and NNMF model performance was not statistically different from using no dilation at all. The effect of over-dilating was further exacerbated when 90% of ratings were observed, with NNMF performing worse compared to no dilation $b$ = 2.2% [1.2% 3.3%], $p$ < .001 (Fig 5 right panel, right group, pink bar).

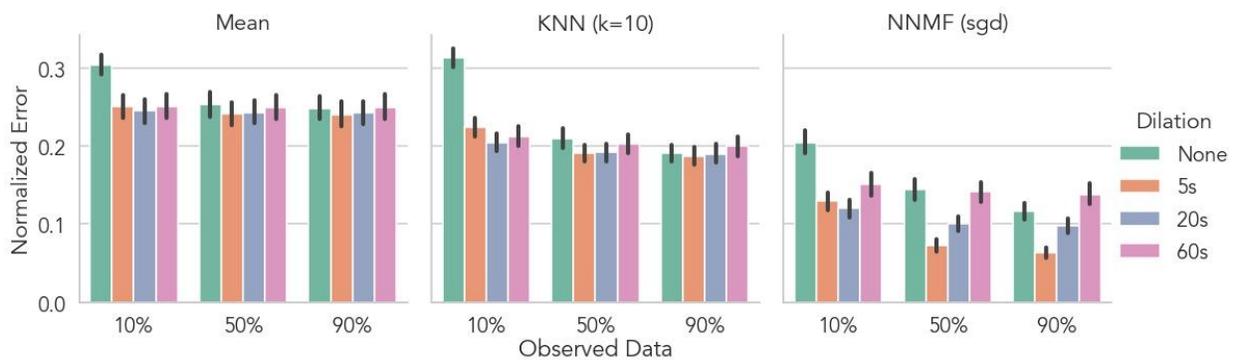

**Figure 5 | Effect of temporal dilation on emotional motional video rating recovery.** Each bar depicts the cross-validated normalized RMSE averaged across users and videos for a given level of temporal

dilation and data sparsity. Error bars depict 95% bootstrapped confidence intervals over users. Lower values indicate better model performance and can be interpreted as the percentage error on the original scale of the data. For all algorithms, when very few ratings were observed (10% of data), dilation substantially improved model predictions. However, the benefit of dilation was reduced as more data were observed. Additionally, when at least 50% of observations were observed, both CF models were sensitive to over-dilation (i.e. kernel width >= 20s). Dilation kernel widths should therefore be chosen as a function of the dynamics of emotional states elicited by stimuli as well as the sparsity of the dataset.

Taken together, these results suggest that CF can be an effective approach for recovering user ratings with large individual differences elicited by naturalistic dynamic stimuli [65]. NNMF trained via SGD along with a small amount of temporal dilation proved to be the most effective algorithm tested at every level of sparsity and even outperformed MICE, a popular multivariate imputation technique. Even with only 10% of observed ratings, 5s of dilation can recover ratings with a small amount of error, approaching performance equivalent to having observed 90% of ratings with no dilation (Fig 6). In practice, researchers may want to tune the amount of dilation employed as a function of the sample rate of data collection as well as the temporal dynamics of the emotional states elicited by the stimuli and intrinsic auto-correlation of the underlying user experience being reported (e.g. moment-to-moment changes in subjective feelings).

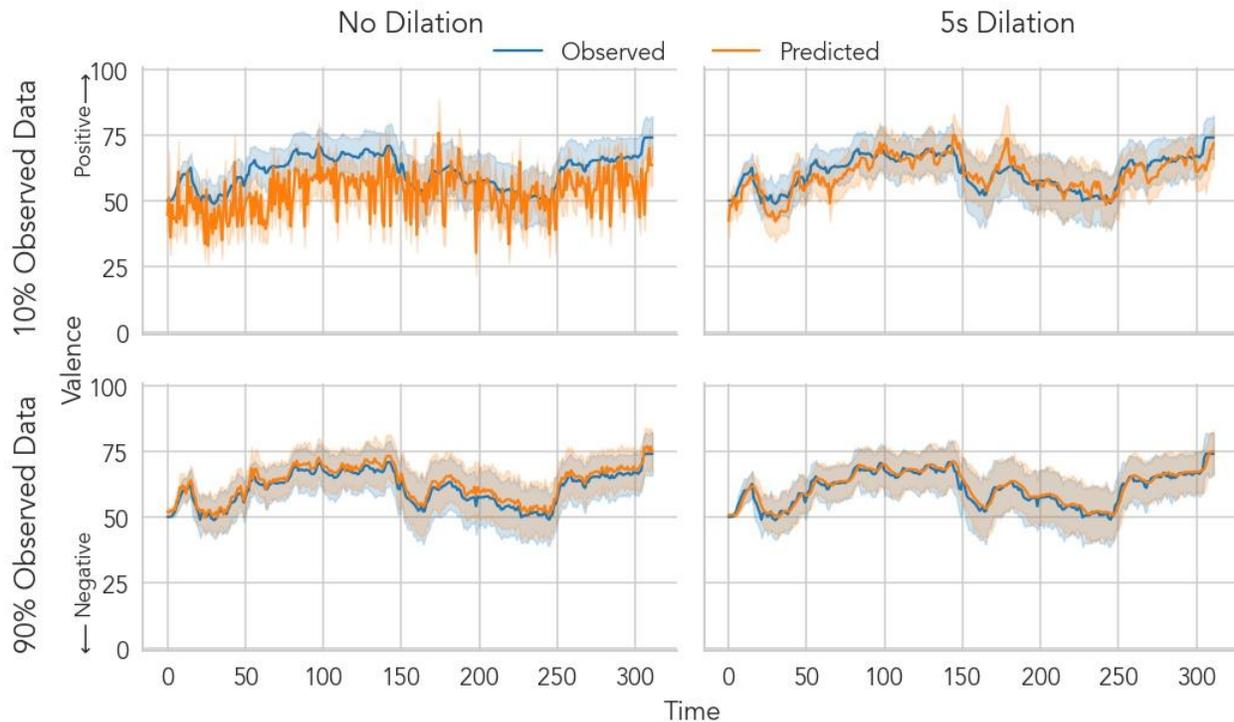

**Figure 6 | Illustrative example of temporal dilation on emotional motional video rating recovery.** An illustrative example of training a NNMF collaborative filtering model using stochastic gradient descent with varying levels of sparsity and dilation. A model trained with 10% data and no dilation performs poorly at reconstructing time-series ratings (top left). However, with a small amount of dilation (5s) prior to training (top right), predictions are almost on par with a model trained on 90% data and no dilation (bottom left). Dilation only offers marginal improvement when sparsity is low (bottom right). Blue lines depict the ground

truth mean time-series rating to a single emotional video clip. Orange lines depict reconstructed time-series using model predictions. Error bands depict 95% bootstrapped confidence intervals.

## 3.4 Recovering Social Decisions

The two previous studies focused on recovering self-reported ratings of subjective emotional experiences, however, CF should in principle be effective with any type of data reflecting stable individual preferences. To test this, in Dataset 3 we explored how well CF could recover individual behavioral decisions when there are known individual variations in social preferences [41,47]. In this experimental context, we found that only CF employed using NNMF recovered behavior significantly better than the Mean model at all levels of sparsity (Fig 7, all *p*s < .001) with reductions in error ranging from 3-4%. The KNN model demonstrated significantly worse performance than the Mean model when less than 30% of ratings were observed (all *p*s < .05) and comparable performance when observed data ranged between 30-60%. Only when at least 60% of ratings were observed did the KNN model outperform the Mean model with reductions in error ranging from 1.6-2% (all *p*s < .001). As in Dataset 2, when the amount of observed data increased, Mean model error rates quickly converged but never improved beyond 15% error, suggesting that the average decision for each item (experimental trial) served as a poor model for individual variability in each user's ratings. This is likely because users comprised various sub-groups driven by different preferences and motivations [41,47]. The MICE model showed significantly worse performance at all levels of sparsity compared to the NNMF model (all *p*s < .01) and significantly worse performance compared to the KNN model when 50% or more data were observed (all *p*s < .01). MICE performance was similar to the Mean model at all levels of sparsity except 10%, 80% and 90% where it performed significantly worse (all *p*s < .01).

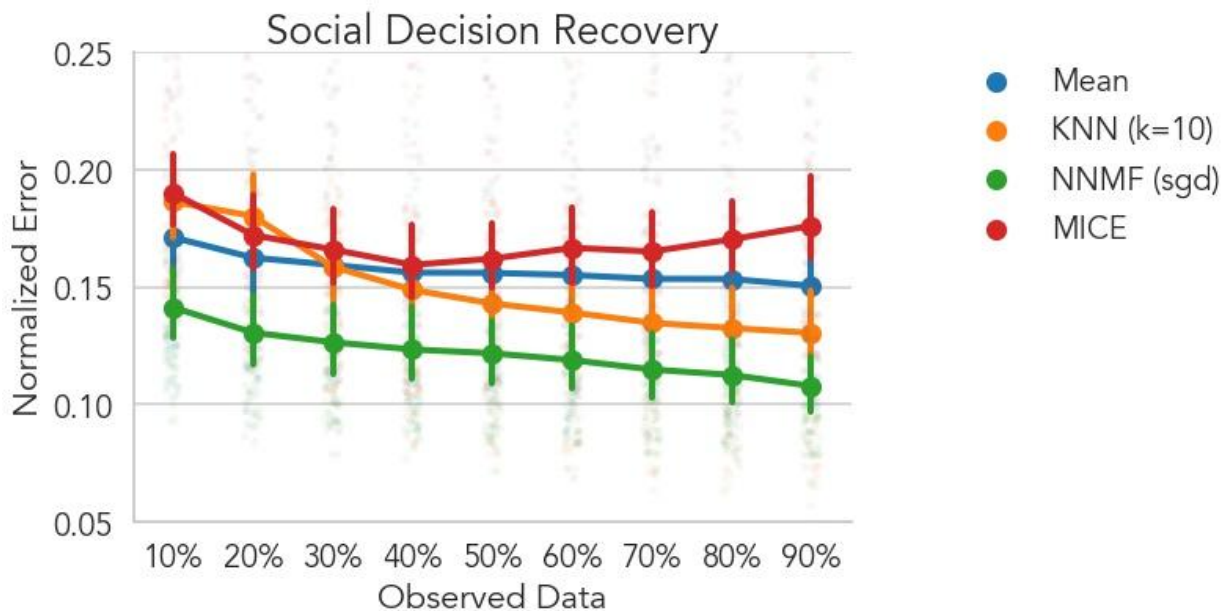

**Figure 7 | Social decision recovery.** Each line depicts the cross-validated normalized RMSE averaged across users. Error bars depict 95% bootstrapped confidence intervals over users and faded points depict mean error for each user. Lower values indicate better model performance and can be interpreted as the percentage error on the original scale of the data. NNMF trained with stochastic gradient descent significantly outperformed the Mean model at every level of dataset sparsity with average error reduction ranging from 3-4%. KNN performance only exceeded the Mean model when at least 60% of data were observed with error reduction ranging between 1.6-2%, likely due to issues computing reliable similarity estimates between users when data were very sparse. MICE performed worse than NNMF at all levels of sparsity and worse than KNN when 50% or more data were observed. For behavioral data with different user sub-groups, CF offers substantial benefits over simpler techniques like mean imputation.

To explore how CF performed separately for user sub-groups, we stratified model performance according to the original moral phenotype labels assigned by van Baar et al [41][3]. Specifically, all users were assigned to separate groups based upon their moral strategy inferred from the authors' computational model (Fig 8). Greedy participants were primarily concerned with maximizing their own financial payoff. Guilt-averse participants were primarily concerned with minimizing their relationship partner's disappointment. Inequity-averse participants were primarily concerned with minimizing discrepancies between payoffs to themselves and their partner. Moral opportunists switched between guilt-averse and inequity-averse strategies in different experimental contexts depending on whichever provided the largest payoff. Only the NNMF model significantly out performed the Mean model for all phenotypes at all levels of data sparsity ranging from 3-14% reduction in error, all $p$s < .05[4]. NNMF also significantly outperformed MICE for all phenotypes at all levels of sparsity, all $p$s < .01. For the greedy phenotype, KNN models performed similarly to the Mean model (Fig 8 top right). For all other phenotypes, KNN models performed similarly or significantly worse than the Mean model when data were sparse, but exceeded Mean model performance as more data were observed: guilt averse - 60%; inequity averse 50%; moral opportunist 80%. KNN likely performed worse than NNMF because sparse observations from any given user were unrepresentative of their overall behavioral strategy thereby making explicit user similarity a foundation generating predictions. For example, with limited data, a subset of one guilt-averse user's choices may have looked more similar to another inequity-averse user's choices rather than more similar to a different guilt-averse user.

# 4. Discussion

## 4.1 Summary of Findings

---

[3] Critically, models were always trained using *all* users without additional stratification based upon moral phenotypes. This more accurately reflects circumstances in which researchers do not know a single user's phenotype *a priori*.
[4] The Bonferroni-corrected p-value was marginal ($p$ = 0.08) when comparing the NNMF to the Mean model at 10% observed data for inequity-averse users.

In the present work, we tested a novel approach to address measurement issues in affective science using a computational modeling approach known as collaborative filtering. By leveraging structured variation across individual experiences we tested how well emotional responses could be inferred from sparsely sampled self-reports. We evaluated the performance of CF in recovering missing data across three common experimental contexts used to study emotion: rating normed static stimuli, continuously rated dynamic stimuli, and decisions made in social contexts.

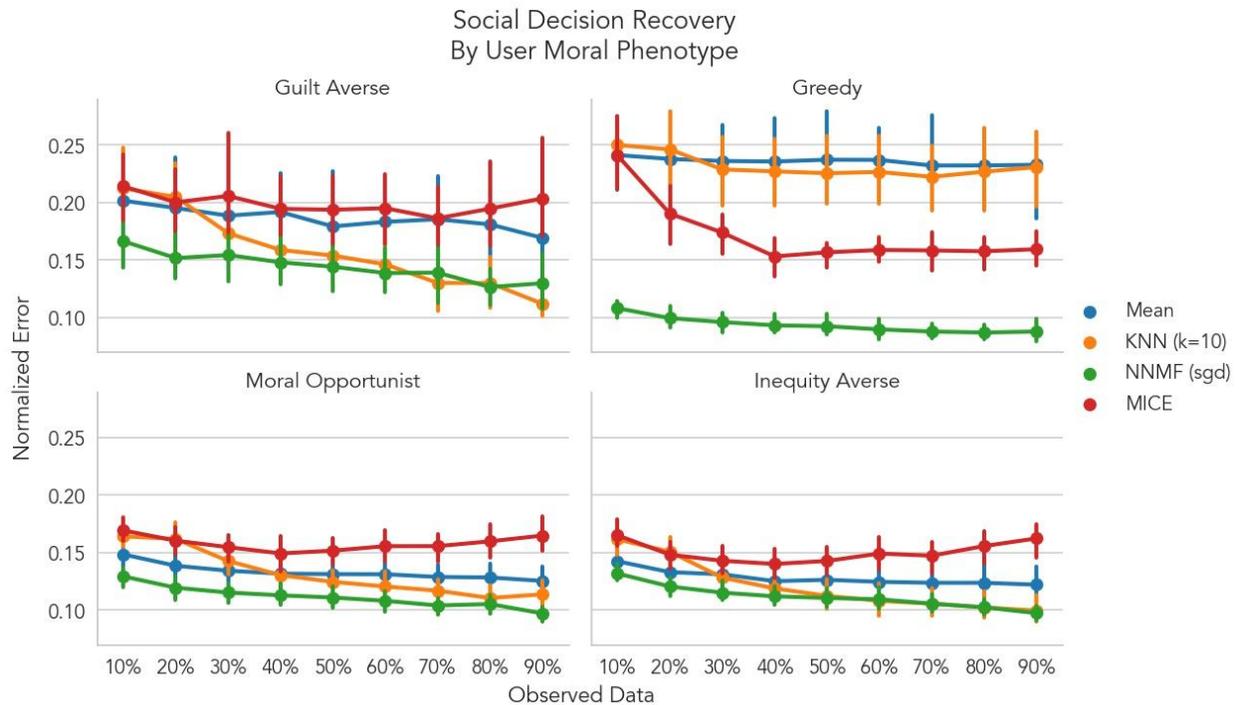

**Figure 8 | Social decision recovery by original user moral phenotype.** Each panel depicts model performance split by the original moral phenotype assigned to each user by Van Baar et al [41]. Each line depicts the cross-validated normalized RMSE averaged across users. Error bars depict 95% bootstrapped confidence intervals. Lower values indicate better model performance and can be interpreted as the percentage error on the original scale of the data. Only NNMF trained with stochastic gradient descent outperformed all other algorithms at every level of dataset sparsity for all moral phenotypes. NNMF also outperformed MICE at all levels of sparsity for all phenotypes. KNN models showed similar performance to mean imputation until a majority of observations were observed: guilt averse - 60%; inequity averse 50%; moral opportunist 80%

Dataset 1 probed emotional responses elicited by static affective image stimuli (IAPS and GAPED [36,40]). This approach is consistent with the status quo in emotion research [5] and served as an opportunity to test whether CF provided any additional benefits above and beyond stimulus norming. Consistent with our predictions, neither CF approach (e.g., KNN or NNMF) provided a substantive benefit over simple mean imputation. This is because the process of norming stimuli inherently minimizes individual differences (Fig 1), thus making the average response to a particular image a sufficient proxy for any single new user's response. While static normed stimuli can provide a straightforward and simple approach for eliciting emotional responses, it does so at the expense of reducing the diversity of emotional experiences and individual variation. In the absence of additional context, individuals may *infer* the semantic meaning of an image and the normative emotional response, but may not directly *experience* the intended elicited emotion themselves. More naturalistic, evocative and dynamic stimuli such as movies, audio clips, etc can provide additional context (e.g. narratives) that are not only more engaging but also more representative of how a wider variety of emotional experiences unfold outside of the laboratory [11,65].

Dataset 2 was collected to explicitly explore the utility of CF in inferring emotional reports to rich dynamic experiences that unfold over time (Moth Radio Hour stories). Participants continuously reported the intensity of their affective responses on a bipolar scale while watching short emotionally engaging autobiographical stories. In this experimental context, we found that CF provided large benefits relative to simple mean imputation. In particular NNMF, a latent factor approach to CF, was extremely effective at predicting time-series ratings while only observing 10% of user responses. Normalized error rates in recovering users' true ratings using this method averaged at least 20% error relative to mean imputation which appeared to asymptote to approximately 30% error in this highly sparse context. In addition, by leveraging the intrinsic auto-correlation of time-series data, we were able to further improve model performance using a kernel approach that provided data augmentation by dilating observed responses prior to model fitting. In all cases, *some* amount of temporal dilation always improved model performance relative to no dilation at all. However, CF models experienced a larger boost in performance relative to mean imputation. For example, using only 10% of the data, NNMF with 5-20s of temporal dilation recovered ratings with as little as 7% error, which was about 3.5x better than mean imputation (approximately 25% error). KNN models also proved more effective than mean imputation, but required at least 30% of data to be observed and never exceeded NNMF performance at any level of data sparsity. This is likely because KNN models require explicit representations of user-user similarity which are used to weight other users' ratings when generating predictions for an unobserved user. When data are highly sparse, the reliability of these similarity scores can be poor or inestimable. Together, these findings suggest that CF is highly effective at recovering emotional responses evoked by naturalistic stimuli comprised of longer, more engaging narrative contexts. One potential explanation for the effectiveness of CF in this experimental context is that naturalistic stimuli elicit greater variations in individual emotional experiences.

We also sought to more systematically evaluate the effectiveness of CF in a dataset with known variations in individual phenotypes [47]. In Dataset 3, participants made decisions about how much money to return to investors in the context of a Hidden Multiplier Trust Game. In the original paper, a computational model was used to estimate each individual's social preference, which motivated their decisions (e.g. guilt-aversion, inequity-aversion, greed, and moral-opportunism). Overall, we found that NNMF models were highly effective at recovering user decisions at all levels of sparsity and equally effective regardless of each user's intrinsic motivation (Fig 7). Even when only 10% of users' decisions were observed, error rates ranged between 10-16%, whereas mean imputation error rates ranged from 14-24%. Critically however, mean imputation along with CF employed via KNN were both highly sensitive to the number of users of each motivational phenotype and fared poorly when a phenotype group consisted of fewer than 10 users (greedy and guilt averse, Fig 8). As observed in Dataset 2, this was likely because of the reduced reliability of user-user similarity scores in highly sparse data scenarios. These findings demonstrate that CF can be a valuable tool for modeling more than just self-reported affect, such as social decisions with real financial outcomes.

While CF has been commonly employed as part of recommender systems for predicting user responses to large catalogs of items (e.g. Netflix shows, Amazon items, Spotify songs) [38], our work demonstrates that CF can be equally effective for affective science. Because it can be difficult to assess real-world model performance without directly inspecting the predictions made by a large recommender, we relied on computing normalized RMSE and compared each CF algorithm to simple mean imputation. We note that the best performance observed on our datasets is on par with top performance observed in the famous Netflix Prize challenge. Archived leaderboards report final RMSEs in the range 0.867 - 0.857 [67], while Netflix's own Cinematch algorithm performed with an RMSE of 0.9525 [66]. These performance estimates come from more sophisticated ensemble approaches to CF trained on approximately 100 million ratings and tested on 3 million unobserved ratings. When normalized on the 5pt scale of ratings of the dataset, this amounts to a normalized error rate of 16.9-17.3% similar to observed error rates in all three datasets we tested.

## 4.2 Comparisons to other imputation approaches and practical utility

A key difference between our work and other techniques for imputing missing data (e.g. MICE [42]), is that CF leverages the similarity between *individuals* rather than the similarity between items to generate predictions (imputations). For example, MICE fits a series of multivariate models (e.g. linear regression) to predict each item in a dataset from all other items, and then uses these models to impute missing data. This allows researchers to leverage the observed covariance between different items to make predictions for missing observations. A key assumption underlying this approach is that any variability across individuals for the same item is essentially noise or error around a singular underlying psychometric measure (e.g. valence or arousal). On the other hand, CF takes an entirely different approach and assumes that any variability between individuals reflects *real individual differences* in participants' experiences. As demonstrated in the current work, this variability can be leveraged to predict missing data from one individual using a weighted combination of observations from other individuals (KNN) or by projecting an individual into a latent space constructed from other individuals (NNMF). In other words, where as common MI techniques answer the question: "can we leverage the population average response to items A and B to infer the population average response to item C?" CF techniques answer the question: "can we leverage the similarity between responses made by individuals X, Y, and Z to item A, to infer individual X's response to item B?"

When applied to the datasets in the present work, MICE performed far more inconsistently relative to the NNMF version of CF. In the affective image and social decision datasets, MICE performed significantly worse at all levels of sparsity (Figs 3, 7). In the emotional video dataset, MICE performed worse or similar to NNMF with no dilation kernel when sparsity was high (20-60% observed data) and slightly better when sparsity was low (70-90% observed data). This is likely because the MICE approach could capitalize on inherent auto-correlation in the data, as individual predictors/features in each estimated regression were different time-points. However, when even a small dilation kernel (e.g. 5s) was used with CF, this difference in performance was no longer observed, as the NNMF model outperformed MICE at every level of sparsity (Fig 4). Further, in a dataset with known phenotypic variability (Dataset 3), MICE performed worse than NNMF at all levels of sparsity (Fig 7), likely because the underlying assumption of a *single* average population response to each item was violated. Critically, because MICE leverages item-covariance rather an user-covariance, it was mathematically impossible to impute ratings in the presence of extreme sparsity (60-90% missing data per individual) making it an impractical technique for such experimental situations (Fig S3). Notably, MICE estimation was 2.5-4.5x *slower* than NNMF for the same dataset and sparsity level, and this computational overhead increased with the number of total items in the dataset. For example at 50% sparsity, running MICE on the shortest video in dataset 2 (312 time-points) took 4.48s +/- 0.42s with normalized error of .123 +/- .081  whereas NNMF took only 1.88s +/- .47 with a normalized error of .137 +/- .104. Running MICE on the longest video (1059 time-points) took 19.31s +/- 1.59s with a normalized error of .151 +/- .065 whereas NNMF only took 4.37s +/- 1.07s with a normalized error of .153 +/- .073.

The CF approach can also enable experimenters to efficiently design studies that *deliberately* sparsely-sample participants due to feasibility or psychometric concerns. For example, we have successfully employed CF using an early version of the toolbox in a naturalistic movie watching experiment (Study 4 in [65]). In this work, we collected sparse self-reported ratings of participants' emotions by pausing the movie at random 3-5 minute intervals following uninterrupted viewing periods of 30s - 4.5 minutes and then applied the NNMF CF algorithm with a 60s dilation kernel to infer participants' ratings at *every second of the movie*, thus reconstructing full emotion-timeseries. This allowed us to perform multimodal latent factor analyses that jointly estimated common signals across these inferred ratings, brain responses collected using functional MRI (fMRI), and facial expressions estimated using computer vision techniques [8]. Despite being a highly sparse dataset (> 95% "missing data"), these inferred ratings appeared to accurately reflect participants' emotional experience in a minimally disruptive way. This particular example provides a proof of concept of how sparse sampling techniques can be used to study rich emotional experiences elicited in more naturalistic experimental contexts.

## 4.3 Limitations and Future Work

It is important to note that this work is not without limitations. First, our models were trained on densely sampled self-report which is affected by all the issues we identified throughout this manuscript. CF like all models will be limited by the reliability of the measured training data. Second, while we only tested one dilation kernel shape (box-car) and a limited range of window sizes on Dataset 2. Additional explorations of kernel shapes and sizes could further improve predictive performance and better suit the nature of the data (e.g. slow time-series vs fast time-series data). We also stress that over-dilation is an important issue to be considered. While dilation may appear to improve model performance up to a point, beyond the intrinsic auto-correlation factor of the data itself, dilation smoothes over temporal dynamics resulting in poor performance (Fig 5). Third, while the performance of NNMF models is robust to different sub-groups of individuals (Fig 8), it is possible that on different datasets model performance will be highly sensitive to group size differences. In other words, if sparsity is not approximately uniformly distributed across known sub-groups, CF may fail to make reliable predictions. Altogether, researchers will need to define their error tolerance with respect to their research goals. The use of pilot samples, with densely acquired ratings using the same stimuli and experimental setting could be an effective strategy to determine measurement reliability and expected performance in larger but sparser datasets. Finally, while we report results from a single NNMF algorithm, several variations exist including an early implementation by Lee and Seung [52] that employs a multiplicative update rule rather than stochastic gradient descent to learn user and item factors. We also tested this algorithm and provide an implementation in Neighbors, but note that it performed poorly on these datasets (see Supplementary Materials).

Future work may also examine additional algorithms and parameterizations for extending the CF models tested here. These might include exploring the effect of dimensionality reduction and regularization on model performance and the estimated factor structures. In addition, neighborhood models might explore how well different distance metrics (e.g. cosine, mahalanobis) perform on different datasets beyond the Pearson correlation measure used here. Finally, CF algorithms can be further extended by incorporating additional information about users (often referred to as "implicit feedback") and stimulus content features into "hybrid" models to improve model predictions, especially when collecting additional data to reduce sparsity is not feasible [38].

## 4.4 Conclusion

For decades, researchers have debated the representational structure of emotion, questioning the tradition of discrete emotion categories, and suggesting continuous low-dimensional representations [67,68]. In our view, measurement and elicitation issues are central to this debate. Because researchers have difficulty measuring emotional experiences, they have increasingly relied on normed stimuli, which necessarily reduce the diversity of experiences elicited by the stimuli. These experiences fail to represent the diversity of feelings typically experienced outside the laboratory, which in turn allows subjective ratings to be represented by lower dimensional embeddings such as valence and arousal [33]. Oversimplifying the emotional experiences elicited in the laboratory increases the risk that emotion researchers are fundamentally building a limited science of highly circumscribed image-evoked feelings rather than a broader science of human emotional experiences [34]. This issue has been raised in other domains of research highlighting how the pernicious effects of un-representative designs [35,69] may lead to a cumulative science that fails to generalize beyond the specific stimuli used in the experimental designs [70]. Moreover, a statistical property of undersampling the space of emotional experiences will lead researchers to capitalize on bias error and falsely conclude that a low dimensional space is the best model of the structure of affect [34]. While the use of naturalistic stimuli provides one solution to this issue, it also introduces another - obtaining self-report with high temporal precision. Simultaneously tasking participants with engaging in an experience while also continuously reporting their feelings is likely to introduce additional cognitive effort and costs [29]. This may create a situation in which participants switch between attending to a stimulus and their own internal state, thereby never fully engaging. As a potential solution to this measurement issue, we propose a sparse-sampling strategy, in which participants infrequently report their feelings, thereby providing an opportunity for them to attend more fully with an engaging stimulus, and later using computational modeling to infer feelings at unsampled time-points. We hope this work will inspire future research that employs more novel approaches to the measurement of human emotion and behavior.

## Acknowledgements

This research was supported by funding from the National Institute of Health (R01MH116026 and R56MH080716), and the National Science Foundation (NSF CAREER 1848370).

## Statements and Declarations

The authors have no competing interests to declare that are relevant to the content of this article

## Data Availability

Code and data used to perform the analysis in the paper is available on github: https://github.com/cosanlab/collab_filter. The neighbors Python package is available at https://neighbors.cosanlab.com/.